\documentclass[twocolumn,prl,superscriptaddress,preprintnumbers]{revtex4}[11pt]
\pdfoutput=1
\usepackage{amsmath,amssymb,graphicx} 
\usepackage{epsf,verbatim}
\usepackage{hyperref}
\usepackage{comment}
\usepackage{color}
\usepackage[normalem]{ulem}

\newcommand{\mysection}[1]{{\noindent \bf #1.}}

\newcommand{\abs}[1]{\left| #1 \right|}
\newcommand{\unit}[1]{\mathrm{\,#1}}

\newcommand{\mDM}{m_{\rm DM}}
\newcommand{\sigmae}{\overline{\sigma}_e}
\newcommand{\FDM}{F_{\rm DM}}
\newcommand{\rhoDM}{\rho_{\rm DM}}
\newcommand{\Zeff}{Z_{\rm eff}}
\newcommand{\Eer}{E_{\rm er}}

\newcommand{\simgt}{\mathrel{\lower2.5pt\vbox{\lineskip=0pt\baselineskip=0pt
           \hbox{$>$}\hbox{$\sim$}}}}
\newcommand{\simlt}{\mathrel{\lower2.5pt\vbox{\lineskip=0pt\baselineskip=0pt
           \hbox{$<$}\hbox{$\sim$}}}}

\newcommand{\squishlist}{
 \begin{list}{$\bullet$}
  { \setlength{\itemsep}{0pt}
     \setlength{\parsep}{3pt}
     \setlength{\topsep}{3pt}
     \setlength{\partopsep}{0pt}
     \setlength{\leftmargin}{1.5em}
     \setlength{\labelwidth}{1em}
     \setlength{\labelsep}{0.5em} } }

\newcommand{\squishlisttwo}{
 \begin{list}{$\bullet$}
  { \setlength{\itemsep}{0pt}
     \setlength{\parsep}{0pt}
    \setlength{\topsep}{0pt}
    \setlength{\partopsep}{0pt}
    \setlength{\leftmargin}{2em}
    \setlength{\labelwidth}{1.5em}
    \setlength{\labelsep}{0.5em} } }

\newcommand{\squishend}{
  \end{list}  }

\newcommand{\be}{\begin{equation}}
\newcommand{\ee}{\end{equation}}
\newcommand{\bea}{\begin{eqnarray}}
\newcommand{\eea}{\end{eqnarray}}
\newcommand{\OO}{\mathcal{O}}

\newcommand{\xenon}{XENON10 }

\begin{document}

\title{First Direct Detection Limits on sub-GeV Dark Matter from XENON10} 

\preprint{YITP-SB-01-12}

\author{Rouven Essig}
\thanks{rouven.essig@stonybrook.edu}
\affiliation{C.N.~Yang Institute for Theoretical Physics, Stony Brook University, Stony Brook, NY 11794}
\affiliation{School of Natural Sciences, Institute for Advanced Study, Einstein Drive, Princeton, NJ}

\author{Aaron Manalaysay}
\thanks{aaronm@physik.uzh.ch}
\affiliation{Physics Institute, University of Zurich, Winterthurerstrasse 190, CH-8057,  Zurich, Switzerland}

\author{Jeremy Mardon}
\thanks{jmardon@stanford.edu}
\affiliation{Stanford Institute for Theoretical Physics, Department of Physics, Stanford University, Stanford, CA 94305}

\author{Peter Sorensen}
\thanks{pfs@llnl.gov}
\affiliation{Lawrence Livermore National Laboratory, 7000 East Ave., Livermore, CA 94550, USA} 
 
\author{Tomer Volansky}
\thanks{tomerv@post.tau.ac.il}
\affiliation{Raymond and Beverly Sackler School of Physics and Astronomy, Tel-Aviv University, Tel-Aviv 69978, Israel}

\begin{abstract} 
The first direct detection limits on dark matter in the MeV to GeV mass range  are presented, using XENON10 data. Such light dark matter can scatter with electrons, causing ionization of atoms in a detector target material and leading to single- or few-electron events. 
 We use 15 kg-days of data acquired in 2006 to set limits on the dark-matter--electron scattering cross section. The strongest bound is obtained at 100\,MeV where  $\sigma_e < 3\times 10^{-38} \unit{cm^2}$ at 90\% CL, while dark matter masses between  20\,MeV and 1\,GeV  are bounded by $\sigma_e < 10^{-37}\unit{cm^2}$ at 90\% CL.  This analysis  provides a first proof-of-principle that direct detection experiments can be  sensitive to dark matter candidates with masses well below the GeV scale.
\end{abstract}

\maketitle

 \setcounter{equation}{0} \setcounter{footnote}{0}

\mysection{INTRODUCTION}
\label{sec:intro}
Most current dark matter (DM) direct detection experiments focus on detecting a Weakly Interacting Massive Particle (WIMP) with a mass of $1-1000\,$GeV.  There are two main reasons for this focus.  Theoretically, a WIMP in this mass range can naturally have the correct thermal relic abundance~\cite{BHS:2005}. Experimentally, a WIMP-nucleus scattering event is likely to produce detectable quanta (phonons, scintillation photons, ionization or some combination of these). DM candidates with mass $\lesssim 1\,$GeV typically cannot produce nuclear recoil signals above detector thresholds, and have therefore been largely ignored.   

It is straightforward, however, to theoretically construct well-motivated, viable, and natural DM candidates with sub-GeV masses (e.g.~\cite{Essig:2011nj,Lin:2011gj,Feng:2008ya,Chu:2011be,Graham:2012su}).  Given the current lack of firm experimental evidence for WIMPs in any mass range, it is important to search for other theoretically motivated DM candidates.  
As was recently proposed in~\cite{Essig:2011nj}, sub-GeV DM can lead to observable signals if it scatters with atomic \emph{electrons}, as opposed to nuclei.  This scattering can ionize atoms in a target material, resulting in single-electron signals. As discussed below, few-electron signals may result if the primary ionized electron or de-excitation photons lead to further ionization. 

Dual-phase liquid xenon detectors have demonstrated sensitivity to such small ionization signals \cite{2008edwards,Santos:2011ju,Aprile:2010bt}. In this letter, we present the first direct detection limits on MeV--GeV-mass DM, using 15\,kg-days of exposure of the \xenon experiment obtained with a single-electron trigger threshold~\cite{Angle:2011th}.  We consider the observed rate of one-, two-, and three-electron events.  The origin of these events is unclear, and they are likely to result from background processes. The data nevertheless allow robust limits to be set for DM as light as a few MeV.  

\mysection{DATA SAMPLE} \label{sec:xenon10}
The \xenon Collaboration has reported results from a 12.5 live-day search for
 scattering of low-mass (few-GeV rather than sub-GeV) WIMPs with xenon nuclei~\cite{Angle:2011th}. Particle interactions in the liquid xenon target can produce both ions (Xe$^+$) and excited atoms (Xe$^*$). A fraction of the ions recombine to form 
other Xe$^*$, whose de-excitation process produces 7\,eV scintillation photons. 
Electrons that escape recombination are accelerated away from the interaction site by an electric field, and   
extracted from the liquid to the gas with an efficiency that is essentially unity \cite{1982gushchin,2004aprile}.  
Under the influence of a high electric field in the gaseous xenon ($\sim$10\,kV/cm), each extracted electron produces  
$\mathcal{O}(100)$ scintillation photons \cite{1999bolozdynya}. 
The detector's array of photomultiplier tubes measures an average of 27 of these photoelectrons per extracted electron.

\begin{figure}[t]
\begin{center}
\vskip -0.1cm
\includegraphics[width=0.48\textwidth]{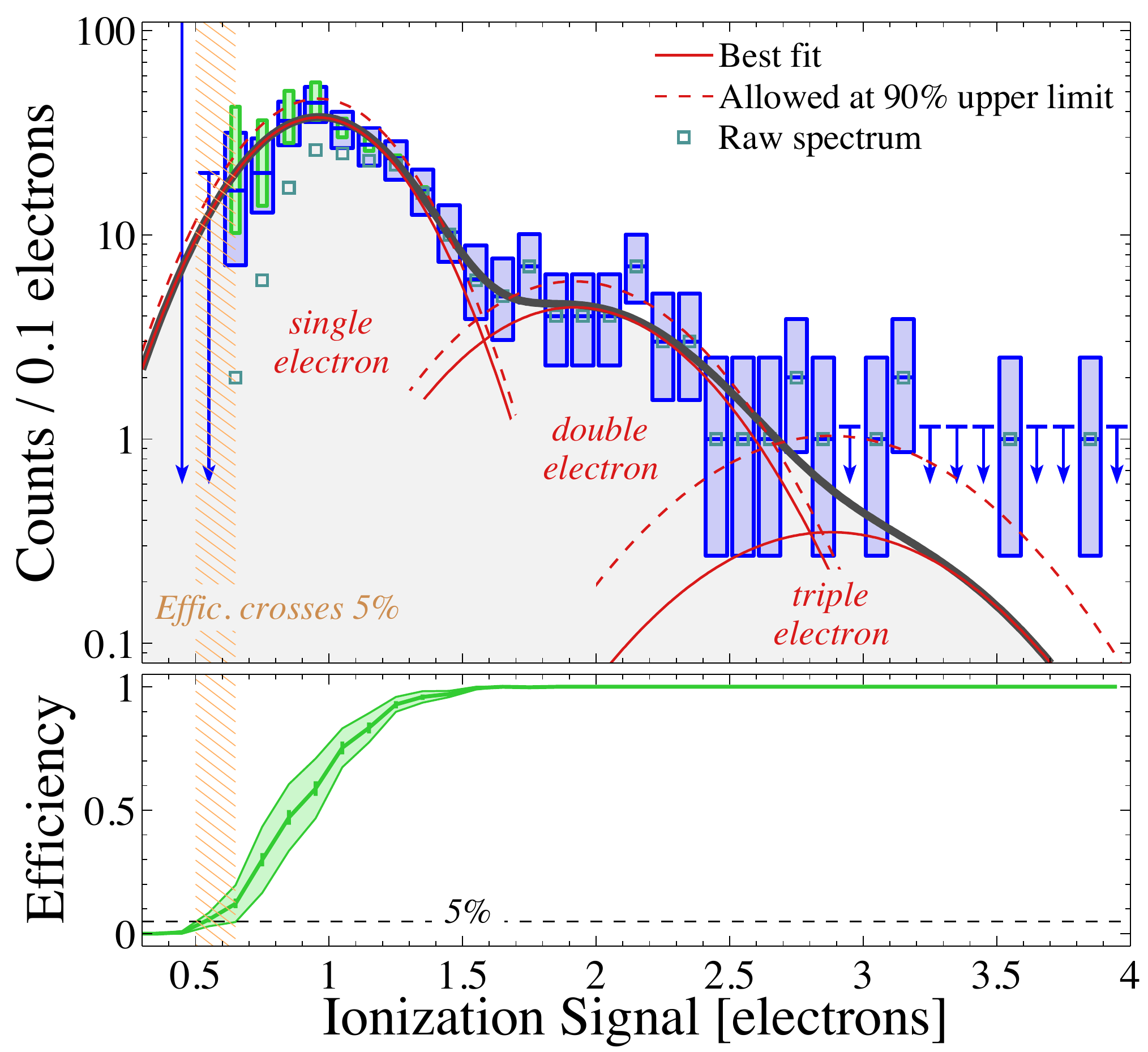}
\vskip -0.1cm
\caption{ {\bf Top}: The spectrum of XENON10 dark matter search data, corrected for trigger efficiency.  Blue boxes indicate statistical uncertainty, while green boxes indicate the systematic uncertainty arising from the the trigger efficiency.  
The efficiency curve crosses 5\% within the orange-hatched vertical band.   
The thick gray curve is the best-fit triple Gaussian function. Thin solid red curves indicated the best-fit individual components. Dashed lines indicate curves allowed at the 90\% upper limit for each component. 
Small open squares indicate the raw spectrum (uncorrected for trigger efficiency) from~\cite{Angle:2011th}. 
Arrows indicate 1-$\sigma$ upper limits on the number of events for bins with no events.  
{\bf Bottom}: The trigger efficiency as determined by Monte Carlo simulation, whose range is chosen such that the efficiency curve crosses 5\% at, or before, the first non-zero bin in the blue histogram.}
\vskip -0.8cm
\label{fig:correctedSpectrum}
\end{center}
\end{figure}

The search for few-GeV dark matter reported in \cite{Angle:2011th} 
imposed a software ionization threshold of 5~electrons due to uncertainties in the ionization yield from very low-energy \emph{nuclear} recoils. However, the \xenon hardware trigger (described in \cite{Aprile:2010bt}) was sensitive to single electrons. Among liquid xenon targets, this is the lowest trigger threshold in all reported dark matter search data. However, the precise trigger efficiency for this data sample was not reported. To better understand the trigger efficiency, we have performed a detailed Monte Carlo simulation of the trigger response to single to few-electron events in the XENON10 detector, based on the information given in Sec.~2.8 of \cite{Aprile:2010bt}. This simulation allows the hardware trigger efficiency to be calculated, and the result is shown in Fig.~\ref{fig:correctedSpectrum} (\emph{bottom}).

The performance and accuracy of our trigger efficiency simulation has been verified by comparing its prediction to the observed trigger roll-off of the calibration spectrum reported in \cite{Sorensen:2010hv}, whose conditions differed from the present data only in the hardware threshold set point.  The good agreement in this known case confirms the validity of the simulation, which is then left with a single free parameter: the hardware threshold set point.  We constrain this threshold by noting that the trigger efficiency curve must ``turn-on'' at, or prior to, the first non-zero bin in the measured spectrum of triggering events, shown in Fig.~2 of~\cite{Angle:2011th}.  In this context, we define the turn-on point as the location where the efficiency curve crosses 5\%, which is indicated by the orange-hatched vertical band in Fig.~\ref{fig:correctedSpectrum}.  If the efficiency were to turn on at a higher point, the peak of the single-electron distribution would be shifted to values much lower than that of the known detector response to these events, demonstrated by Fig.~2 (\emph{top}) of~\cite{Angle:2011th}.

The measured spectrum of triggering ionization events, which we analyze for a signal, is given in Fig.~2 (\emph{top}) of~\cite{Angle:2011th}.  We reproduce this spectrum in Fig.~\ref{fig:correctedSpectrum} (\emph{top}), corrected for the trigger efficiency.  Wide (blue) bars represent statistical uncertainty, while the narrow (green) bars indicate the systematic uncertainty introduced by the range of allowed trigger efficiencies.  This spectrum is fit by a triple Gaussian function with five free parameters: the heights, $H_i$, of the three components and the mean and width of the first component ($\mu_1$, $\sigma_1$).  The means, $\mu_i$, and widths, $\sigma_i$, are constrained to follow the relations $\mu_i = \mu_1i$ and $\sigma_i = \sigma_1\sqrt{i}$, respectively, where
$i=1,2,3$ identifies the Gaussian component.  Individual marginal posterior probability distributions are obtained for the event rates of the three components, $r_i = H_i\sigma_i\sqrt{2\pi}/\epsilon S\Delta x$, 
where $\epsilon=0.92$ is the overall cut efficiency reported in \cite{Angle:2011th}, $S$\,=\,15\,kg-days is the exposure, and $\Delta x$\,=\,0.1\,electrons is the histogram bin width.  From these, upper limits are extracted taking the measured spectrum to be due entirely to signal (\emph{i.e.}~no background subtraction).  The result of the fit, including statistical and systematic uncertainties, gives 90\% upper confidence bounds of 
$r_1<23.4$, $r_2<4.23$, and $r_3<0.90$ cts~kg$^{-1}$\,day$^{-1}$.

\mysection{DIRECT DETECTION RATES}
We assume that DM particles scatter through direct interactions with atomic electrons. 
If the DM--electron interaction is independent of the momentum transfer, $q$, then it is completely parametrized by the elastic 
cross section, $\sigma_e$, of DM scattering with a free electron.  For $q$-dependent interactions, we define a cross section 
$\sigmae$ by fixing $q=\alpha m_e$ in the matrix element~\cite{Essig:2011nj}. The $q$ dependence of the matrix element is then described by a DM form-factor, $\FDM(q)$; for example, if the interaction proceeds through a massless vector mediator then $\FDM=(\alpha m_e/q)^2$.

A large fraction of the kinetic energy carried by a DM particle, $E_{\rm DM} = m_{\rm DM} v^2/2 \simeq 10\unit{eV} (m_{\rm DM}/20 \unit{MeV})$, can be transferred to a primary ionized electron. We treat the target electrons as single-particle states bound in isolated xenon atoms, using the numerical RHF bound wavefunctions tabulated in~\cite{Bunge:1993}. 
The electron recoils with energy $\Eer$, with a differential ionization rate~\cite{Essig:2011nj} 
\begin{equation}
\label{eq:ionization-rate}
\frac{dR_{ion}}{d\ln \Eer}
=  
N_T \frac{\rhoDM}{\mDM} \sum_{n l} \frac{d\langle \sigma_{ion}^{n l} v \rangle}{d\ln \Eer} \, ,
\end{equation}
where $N_T$ is the number of target atoms, $\rhoDM = 0.4\unit{GeV\,cm^{-3}}$ is the local DM density, and the velocity-averaged differential ionization cross section for electrons in the $(n,l)$ shell is given by
\begin{equation}
\label{eq:ionization-cross-section}
\!\!\! \frac{d \langle \sigma_{ion}^{n l} v \rangle}{d\ln{\Eer}} \!=\! \frac{\overline{\sigma}_e}{8 \mu_{\chi e}^2} \!\int\!\!\! q \big|f_{ion}^{n l}(k',q)\big|^2 \big|F_\text{DM}(q)\big|^2 \eta (v_\text{\rm min}) \, d q\, . \!\!
\end{equation}
Here $v_{min} = (|E_{\rm binding}^{n l}| + \Eer)/q +q/2 \mDM$, and
 $\eta(v_{\rm min}\!)$ has its usual meaning $\langle \frac{1}{v} \theta(v\!\!-\!\!v_{\rm min}\!) \rangle$.
We assume a standard Maxwell-Boltzmann velocity distribution with circular velocity $v_0=220$\,km\,s$^{-1}$ and a hard cutoff at $v_{\rm esc}=544$\,km\,s$^{-1}$~\cite{astro-ph/0611671}.

With full shells, the form-factor for ionization of an electron in the $(n, l)$ shell, escaping with momentum $k' = \sqrt{2 m_e \Eer}$ after receiving a momentum transfer $q$, can be written as
\begin{multline}
\label{eq:ionization-form-factor}
\!\! \big|f_{ion}^{n l}(k',q)\big|^2 \!=\!
\frac{4 k'^3}{(2\pi)^3} \!\sum_{l'\, L} (2l+1)(2l'+1)(2L+1) \!
 \begin{bmatrix}
	l & l' & L \\
	0 & 0 & 0
 \end{bmatrix}^2 \\
\times \abs{\int\! r^2 d r \, R_{k' l'}(r) R_{n l}(r) j_L(q r)}^2 \, ,
\end{multline}
where $[ \cdots ]$ is the Wigner 3-$j$ symbol and $j_L$ is a spherical Bessel function. 
The radial wavefunctions $R_{k' l'}(r)$ of outgoing electrons are found by numerically solving the radial Schr\"odinger equation with a central potential $\Zeff(r)/r$. $\Zeff(r)$ is determined from the initial electron wavefunction, assuming it to be a bound state of the same central potential. 
We evaluate the form-factors numerically, cutting off the sum at large $l', L$ once it converges. Only the ionization rates of the 3 outermost shells (5p, 5s, and 4d, with binding energies of 12.4, 25.7, and 75.6 eV, respectively) are found to be relevant.

The energy transferred to the primary ionized electron by the initial scattering process is ultimately distributed into a number 
of (observable) electrons,  $n_e$, (unobserved) scintillation photons, $n_\gamma$, and heat. To calculate 
$n_e$, we use a probabilistic model based on a combined theoretical and empirical understanding of the electron yield of higher-energy electronic recoils.  Absorption of the primary electron energy creates a number of ions, $N_i$, and a number of excited atoms, $N_{\mathrm{ex}}$, whose initial ratio is  determined  to be $N_{\mathrm{ex}}/N_i$\,$\approx$\,$0.2$ over a wide range of energies above a keV~\cite{2007aprile,2002doke}.    
Electron--ion recombination appears well-described by a modified Thomas-Imel recombination 
model~\cite{2009dahl, 2011sorensen}, which suggests that the fraction of ions that recombine, $f_R$, is essentially 
zero at low energy, resulting in $n_e = N_i$ and $n_\gamma = N_{\rm ex}$.
The fraction, $f_e$, of initial quanta observed as electrons is therefore given by $f_e =(1-f_R) (1+N_{\mathrm{ex}}/N_i)^{-1}\approx 0.83$~\cite{2011sorensen}.  The total number of quanta, $n$, is observed to behave, at higher energy, as $n = E_{\mathrm{er}}/W$, where $E_{\mathrm{er}}$ is the outgoing energy of the initial scattered electron and $W=13.8$\,eV is the average energy required to create a single quanta~\cite{2007shutt}.  As with $f_R$ and $N_{\mathrm{ex}}/N_i$,  $W$  is only well measured at energies higher than those of interest to us, and thus adds to the theoretical uncertainty in the predicted rates.  
We use $N_{\rm ex}/N_i = 0.2$, $f_R=0$ and $W=13.8$\,eV to give central limits,
and to illustrate the uncertainty we scan over the ranges $0<f_R<0.2$, $0.1<N_{\mathrm{ex}}/N_i<0.3$, and 
$12.4<W<16$\,eV. 
The chosen ranges for $W$ and $N_{\mathrm{ex}}/N_i$ are reasonable considering the available data \cite{2002doke,Aprile:2010bt,2007aprile,2001RaPC...60..291S}. The chosen range for $f_R$ is conservative considering the fit of the Thomas-Imel model to low-energy electron-recoil data \cite{2009dahl}.

\begin{figure}[t!]
\begin{center}
\includegraphics[width=0.48\textwidth]{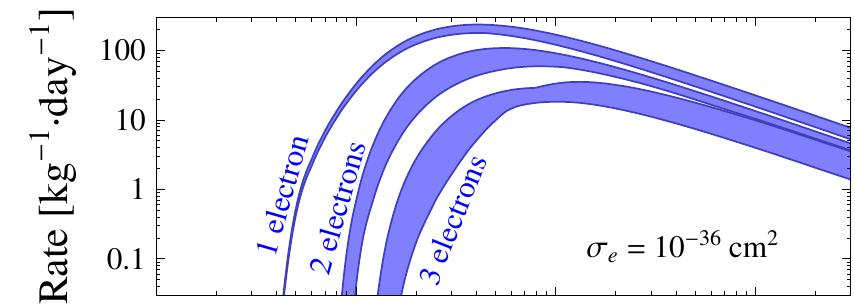}
\vskip -1pt
\includegraphics[width=0.48\textwidth]{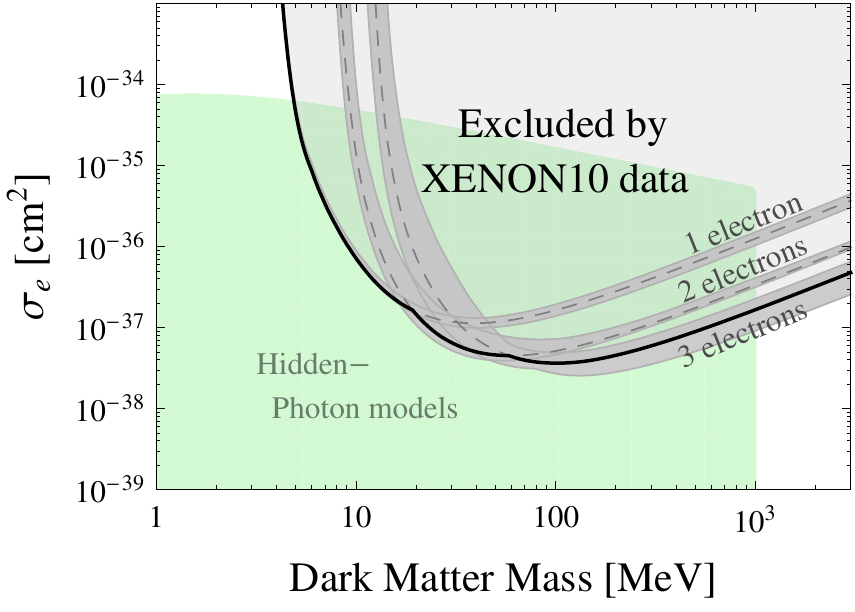}
\vskip -0.2cm
\caption{
{\bf Top}: Expected signal rates for 1-, 2-, and 3-electron events for a DM candidate with $\sigma_e = 10^{-36} \unit{cm^2}$ and $\FDM = 1$.
Widths indicate theoretical uncertainty (see text).
{\bf Bottom}: 90\% CL limit on the DM--electron scattering cross section $\sigma_e$ (black line). Here the interaction is assumed to be independent of momentum transfer ($\FDM = 1$). The dashed lines show the individual limits set by the number of events in which 1, 2, or 3 electrons were observed in the \xenon data set, with gray bands indicating the theoretical uncertainty.
The light green region indicates the previously allowed parameter space for DM coupled through a massive hidden photon (taken from~\cite{Essig:2011nj}).
}
\vskip -0.9cm
\label{fig:result}
\end{center} 
\end{figure}

We extend this model to DM-induced ionization as follows.
We calculate the differential single-electron ionization rate following Eqs.~(\ref{eq:ionization-rate}--\ref{eq:ionization-form-factor}). 
We assume the scattering of this primary electron creates a further
$n^{(1)}={\rm Floor}(\mbox{E}_{er}/W)$ quanta.
In addition, for ionization of the next-to-outer 5s and 
4d shells, we assume that the photon associated with the de-excitation of the 5p-shell electron, with energy $13.3$ or $63.1\,$eV, 
can photoionize, creating another $n^{(2)} = 0$ (1) or 4 quanta, respectively, for $W> 13.3$ eV ($<13.3$ eV). 
 The total number of detected electrons is thus $n_e = n_e' + n_e''$, 
 where $n_e'$ represents the primary electron and is thus 0 or 1 with probability $f_R$ or $(1-f_R)$, respectively,
and $n_e''$ follows a binomial distribution with $n^{(1)}+n^{(2)}$ trials and success probability $f_e$. 
This procedure is intended to reasonably approximate the detailed microscopic scattering processes, 
but presents another $\mathcal{O}(1)$ source of theoretical uncertainty. 
The 1-, 2-, and 3-electron rates as a function of DM mass for a fixed cross section and $\FDM=1$ are shown in Fig.~\ref{fig:result} (\emph{top}).  
 The width of the bands arises from scanning over  $f_R$, $N_{\mathrm{ex}}/N_i$ and $W$, as described above, and illustrates the theoretical uncertainty.

\begin{figure}[t!]
\begin{center}
\includegraphics[width=0.48\textwidth]{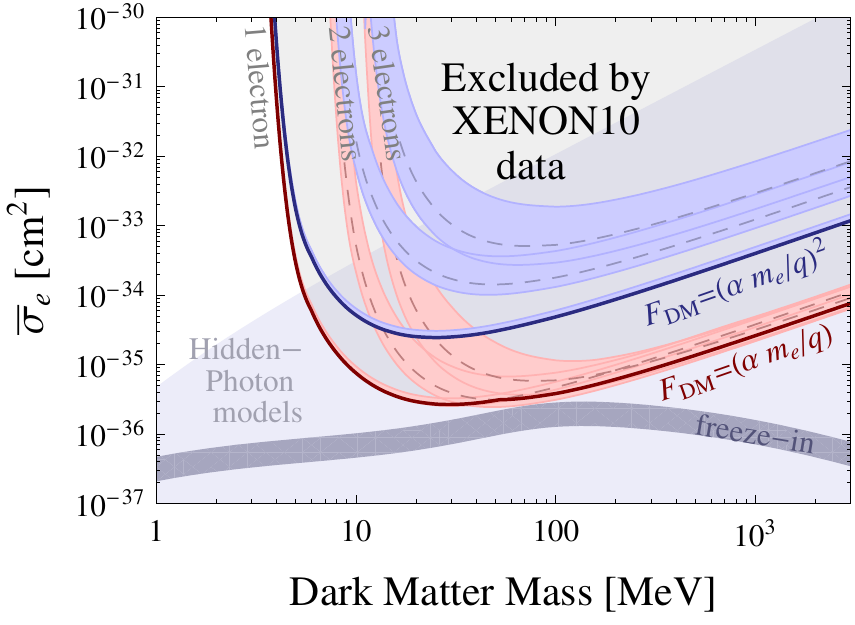}
\vskip -0.2cm
\caption{90\% CL exclusion limits on $\sigmae$ for candidates with a DM form-factor $\FDM(q)$ of $(\alpha m_e/q)$ (red/lower line), corresponding to DM with an electric dipole moment, and $(\alpha m_e/q)^2$ (blue/upper line), corresponding to DM scattering through a very light mediator.
Dashed lines and bands are as in Fig.~\ref{fig:result}.
The pale blue region shows the previously allowed parameter space for DM coupled through a very light hidden photon ($\FDM=(\alpha m_e/q)^2$), with the gray strip indicating the ``freeze-in'' region (taken from~\cite{Essig:2011nj}).
}
\vskip -0.8cm
\label{fig:result-formfactor}
\end{center}
\end{figure}

\mysection{RESULTS} \label{sec:results}
Fig.~\ref{fig:result} (\emph{bottom}) 
shows the exclusion limit in the $\mDM$-$\sigma_e$ plane based on the upper limits for 1-, 2-, and 3-electrons rates in the XENON10 data set (dashed lines), and the central limit (black line), corresponding to the best limit at each mass. 
The gray bands show the theoretical uncertainty, as described above.
This bound applies to DM candidates whose non-relativistic interaction with electrons is momentum-transfer independent ($\FDM=1$). For DM masses larger than $\sim$15\,MeV, the bound is dominated by events with 2 or 3 electrons, due to the small number of such events observed in the data set. For smaller masses, the energy available is insufficient to ionize multiple electrons, and the bound is set by the number of single-electron events.
The light green shaded region shows the parameter space spanned by models in which the DM candidate is a fermion coupled to the visible sector through a kinetically mixed ``hidden photon'' with $\OO$(MeV-GeV) mass, and satisfying all previously known  constraints (from~\cite{Essig:2011nj}; see also~\cite{Lin:2011gj}
\footnote{The allowed region given in~\cite{Lin:2011gj} excludes a small portion in the lower left of the region shown here, due to insufficient self-annihilation of the symmetric DM component. We include this region to allow for more general hidden sectors in which DM has alternate annihilation channels.
}). 

Fig.~\ref{fig:result-formfactor} shows the exclusion limits in the $\mDM$-$\sigmae$ plane for DM candidates whose interaction with electrons is enhanced at small momentum-transfers by a DM form-factor, $\FDM$. 
 The red (lower) curves correspond to $\FDM = (\alpha m_e/q)$, or DM scattering through an electric dipole 
 moment, and the blue (upper) curves to $\FDM = (\alpha m_e/q)^2$, or DM scattering though a very light ($\ll$ keV) scalar or vector mediator.
 Bounds set by 1-, 2-, and 3-electron rates are shown by dashed lines, and the central limits by dark lines. The bands illustrate the theoretical uncertainty. Both form-factors suppress the relative rate of events with larger energy deposition, and so reduce the fraction (and hence the importance) of events containing multiple electrons.
 The pale blue region shows the parameter space for DM coupled through a very light hidden-photon mediator, and satisfying all previously known constraints, with the gray strip showing where the correct abundance is achieved through ``freeze-in" (from~\cite{Essig:2011nj}).
These regions should be compared to the blue exclusion curve.

\mysection{DISCUSSION} \label{sec:summary}
The results above demonstrate, for the first time, the ability of direct detection experiments to probe 
DM masses far below a GeV.
It is encouraging that with only 15\,kg-days of data, 
 and no attempt to control single-electron backgrounds,
the \xenon experiment 
 places meaningful bounds down to masses of a few MeV.

 It should be emphasized that this analysis lacks the ability to distinguish signal from background.  One promising 
 method is the expected annual modulation of the signal.  
As discussed in~\cite{Essig:2011nj},
 additional discrimination may be possible via the collection of individual photons, phonons~\cite{Formaggio:2011jt}, or ions, although at present such technologies have yet to be established.

Independently, this type of search could be significantly improved with a better understanding of few-electron backgrounds.  A quantitative background estimate was not made in~\cite{Angle:2011th},
 making background subtraction impossible.
 Single-electron ionization signals have been studied, and potential causes discussed, by XENON10~\cite{Aprile:2010bt}, ZEPLIN-II~\cite{2008edwards}, and ZEPLIN-III~\cite{Santos:2011ju}.  Possible sources include photo-dissociation of negatively charged impurities, spontaneous emission of electrons that have become trapped in the potential barrier at the liquid-gas interface, and field emission in the region of the cathode.  The former two processes would not be expected to produce true two- or three-electron events, although
single electron events may overlap in time, giving the appearance of an isolated, double-electron event. 
With a dedicated study, these backgrounds could be quantitatively estimated and reduced. 
 
With larger targets and longer exposure times, ongoing and upcoming direct detection experiments such as XENON100, XENON1T, LUX, and CDMS, should be able to improve on the sensitivity reported here.  Such improvements may require optimizations of the triggering thresholds, and will strongly benefit from additional studies of the backgrounds.

\vskip 0.2mm
\begin{center} 

{\bf Acknowledgements}
\end{center}
\vskip -1mm
R.E.~acknowledges support from NSF grant PHY-0969739.  J.M.~is supported by a Simons Postdoctoral Fellowship.  T.V.~is supported in part by a grant from the Israel Science Foundation, the US-Israel Binational Science Foundation and the EU-FP7 Marie Curie, CIG fellowship.

\end{document}